\documentclass[12pt]{article}
\usepackage{a4wide}

\newcommand{\pfrac}[2]{\left(\frac{#1}{#2}\right)}

\begin{document}

\begin{flushright}
MZ-TH/12-15\\
SI-HEP-2012-07\\
1204.0694 [hep-ph]
\end{flushright}

\begin{center}
{\Large\bf A numerical test of differential equations for\\[3pt]
  one- and two-loop sunrise diagrams using\\[12pt]
 configuration space techniques}

\vspace{1truecm}

{\large \bf S.~Groote$^{1,2}$, J.G.~K\"orner$^2$ and
  A.A.~Pivovarov$^{3,4}$}\\[.4truecm]
$^1$F\"u\"usika Instituut, Tartu \"Ulikool,
  T\"ahe 4, EE-51010 Tartu, Estonia\\[.3truecm]
$^2$Institut f\"ur Physik der Johannes-Gutenberg-Universit\"at,\\
  Staudinger Weg 7, D-55099 Mainz, Germany\\[.3truecm]
$^3$Institute for Nuclear Research of the\\
  Russian Academy of Sciences, Moscow 117312, Russia\\[.3truecm]
$^4$Department Physik der Universit\"at Siegen,\\
  Walter-Flex-Str.~3, D-57068 Siegen, Germany
\end{center}

\begin{abstract}
We use configuration space methods to write down one-dimensional
integral representations for one- and two-loop sunrise diagrams
(also called Bessel moments) which we use to numerically check on the 
correctness of the second order differential equations for one- and two-loop 
sunrise diagrams that have recently been discussed in the literature.
\end{abstract}

\newpage

\section{Introduction}
Sunrise-type diagrams have been under investigation since many years. Exact
analytical results can be obtained only for special mass or kinematic
configurations such as for the equal or zero mass cases or for the threshold
region. For example, threshold expansions of the non-degenerate massive
two-loop sunrise diagram have been studied in Refs.~\cite{Davydychev:1999ic,%
Groote:2000kz}. The construction of differential equations for the
corresponding correlator function provides some hope that by solving these
differential equations, a general analytical solution can be obtained.
Recently, mathematical methods were used to construct the coefficients of such
a differential equation in a systematic way~\cite{MullerStach:2011ru}. This
work supplements the work of Kotikov~\cite{Kotikov:1990kg} and Remiddi
{\it et al.\/}~\cite{Remiddi:1997ny,Caffo:1998du,Laporta:2004rb} on the same
subject. While traditionally the correlator is calculated in momentum space,
configuration space techniques allow for a surprisingly simple solution for
sunrise-type diagrams: The correlator in configuration space is just a product
of single propagators which in turn can be expressed by modified Bessel
functions of the second kind. Transforming back to momentum space, one ends up
with a one-dimensional integral over Bessel functions, known as Bessel
moments~\cite{Bailey:2008ib,Broadhurst:2008mx}. As outlined in a series of
papers~\cite{Mendels:wc,Groote:1998ic,Groote:1998wy,Groote:1999cx,%
Groote:2000kz,Delbourgo:2003zi,Groote:2004qq,Groote:2005ay}, the corresponding
one-dimensional integral can be easily integrated numerically for an arbitrary
number of propagators with different masses in any space-time dimension.
Therefore, configuration space techniques can be used to numerically check the
differential equations for the correlator function obtained by other means.
This will be detailed in this note.

The paper is organized as follows: In Sec.~2 we introduce the configuration
space techniques which will be used in Sec.~3 to check the differential
equations for one-loop sunrise-type diagrams. In Sec.~4 we check the
differential equations for the two-loop sunrise diagrams for the equal mass
case, while in Sec.~5 we will deal with nondegenerate masses. Our conclusions
can be found in Sec.~6. Even though the configuration space techniques are
well suited to treat general $D\neq4$ space-time dimensions, we will mainly
deal with the case of $D=2$ space-time dimensions in this paper. For reasons
of simplicity, throughout this paper we work in the Euclidean domain. The
transition to the Minkowskian domain can be obtained as usual by a Wick
rotation (or, equivalently, by replacing $p^2\to-p^2$).

\section{Configuration space techniques}
In configuration space, the $n$-loop $n$-particle irreducible correlation
function
\begin{equation}
\Pi(x)=\langle 0|T\bar j(x)j(0)|0\rangle
\end{equation}
connecting the space-time points $0$ and $x$ is given by the product of the
propagators,
\begin{equation}
\Pi(x)=\prod_{i=1}^{n+1}D(x,m_i),
\end{equation}
where the free propagator of a particle with mass $m$ in $D$-dimensional
(Euclidean) space-time is given by
\begin{equation}\label{prop}
D(x,m)=\frac1{(2\pi)^D}\int\frac{e^{i(p\cdot x)}d^Dp}{p^2+m^2}
  =\frac{(mx)^\lambda K_\lambda(mx)}{(2\pi)^{\lambda+1}x^{2\lambda}}
\end{equation}
($D=2\lambda+2$). $K_\lambda(z)$ is the McDonald function (modified Bessel
function of the second kind). Note that $p$ and $x$ in the integral expression
of Eq.~(\ref{prop}) are $D$-dimensional Lorentz vectors, i.e.\
$p\cdot x=p_\mu x^\mu$, while the quantity $x$ in the rightmost expression of
Eq.~(\ref{prop}) (and, therefore, also in the argument of the propagator) 
denotes the absolute value $x=\sqrt{x_\mu x^\mu}$. In the limit $mx\to 0$ at
fixed $x$, the propagator simplifies to
\begin{equation}
D(x,0)=\frac1{(2\pi)^D}\int\frac{e^{i(p\cdot x)}d^Dp}{p^2}
  =\frac{\Gamma(\lambda)}{4\pi^{\lambda+1}x^{2\lambda}},
\end{equation}
where $\Gamma(\lambda)$ is Euler's Gamma function.

In this note we write the $n$-particle irreducible correlator function in
(Euclidean) momentum space. The momentum space $n$-particle irreducible
correlator function is given by the Fourier transform of the $n$-particle
irreducible correlator function $\Pi(x)$ in configuration space,
\begin{equation}
\label{fourier}
\tilde\Pi(p)=\int\Pi(x)e^{i(p\cdot x)}d^Dx.
\end{equation}
As a product of propagators, $\Pi(x)$ in~(\ref{fourier}) depends only on the
absolute value $x=\sqrt{x_\mu x^\mu}$, Therefore, one proceeds by first
integrating the exponential factor over the $D-1$ dimensional hypersphere. We
write $d^Dx=x^{D-1}d^D\hat x\,dx$ where $d^D\hat x$ denotes the
$D-1$ dimensional integration measure over the $D-1$ dimensional hypersphere.
The integration of the exponential factor over the unit sphere gives
\begin{equation}\label{measure}
\int e^{i(p\cdot x)}d^D\hat x=2\pi^{\lambda+1}\pfrac{px}2^{-\lambda}
  J_\lambda(px).
\end{equation}
$J_\lambda(z)$ is the Bessel function of the first kind. As before, $p$ and
$x$ on the right hand side of Eq.~(\ref{measure}) stand for the absolute
values $p=\sqrt{p_\mu p^\mu}$ and $x=\sqrt{x_\mu x^\mu}$. Therefore, the
correlator in momentum space depends only on the absolute value of the
momentum,
\begin{equation}
\tilde\Pi(p)=2\pi^{\lambda+1}\int_0^\infty\pfrac{px}2^{-\lambda}J_\lambda(px)
  \Pi(x)x^{2\lambda+1}dx.
\end{equation}
This is the central formula for our numerical verification of the correctness
of the differential equations.

\section{The one-loop case}
In Ref.~\cite{Remiddi:1997ny}, Remiddi explains how to obtain the differential
equation for the one-loop sunrise-type diagram with arbitrary masses and
dimensions. By applying the integration-by-parts technique to the correlator in
momentum space, recurrence relations can be obtained. Finally, Euler's theorem
for homogeneous functions connects the loose ends of the iterative steps
involving partial derivative with respect to $p^2$. We have numerically 
checked all these
steps and have found numerical consistency -- up to Stokes' contributions due
to surface terms in integer space-times dimensions.

To be more precise, the integral
\begin{equation}
\int\frac{d^Dk}{(2\pi)^D}\frac\partial{\partial k_\mu}
  \pfrac{v_\mu}{(k^2+m_1^2)\left((p-k)^2+m_2^2\right)}
\end{equation}
for $v=k,p$ (or a linear combination of both) leads to a surface term which can
be assumed to vanish (up to Stokes' contributions). The integration-by-parts
technique consists in calculating the integral explicitly and one then 
expresses the
result in terms of scalar integrals
\begin{equation}
S(\alpha_1,\alpha_2):=\int\frac{d^Dk}{(2\pi)^D}\frac1{(k^2+m_1^2)^{\alpha_1}
  \left((p-k)^2+m_2^2\right)^{\alpha_2}}.
\end{equation}
The (two) resulting recurrence relations read
\begin{eqnarray}
0&=&D\,S(\alpha_1,\alpha_2)+2\alpha_1\left(m_1^2S(\alpha_1+1,\alpha_2)
  -S(\alpha_1,\alpha_2)\right)\nonumber\\&&\strut
  +\alpha_2\left((p^2+m_1^2+m_2^2)S(\alpha_1,\alpha_2+1)
  -S(\alpha_1-1,\alpha_2+1)-S(\alpha_1,\alpha_2)\right),
  \qquad\label{recur1}\\[12pt]
0&=&-\alpha_1\left((p^2-m_1^2+m_2^2)S(\alpha_1+1,\alpha_2)
  -S(\alpha_1+1,\alpha_2-1)+S(\alpha_1,\alpha_2)\right)\nonumber\\&&\strut
  +\alpha_2\left((p^2+m_1^2-m_2^2)S(\alpha_1,\alpha_2+1)
  -S(\alpha_1-1,\alpha_2+1)+S(\alpha_1,\alpha_2)\right).\quad\label{recur3}
\end{eqnarray}
Eq.~(\ref{recur3}) can be replaced by Eq.~(\ref{recur1}) with the two lines
interchanged,
\begin{eqnarray}
0&=&D\,S(\alpha_1,\alpha_2)+2\alpha_2\left(m_2^2S(\alpha_1,\alpha_2+1)
  -S(\alpha_1,\alpha_2)\right)\nonumber\\&&\strut
  +\alpha_1\left((p^2+m_1^2+m_2^2)S(\alpha_1+1,\alpha_2)
  -S(\alpha_1+1,\alpha_2-1)-S(\alpha_1,\alpha_2)\right).\qquad\label{recur2}
\end{eqnarray}
It is obvious that Eq.~(\ref{recur3}) is reproduced as difference of
Eq.~(\ref{recur1}) and Eq.~(\ref{recur2}). Therefore, one has to check only
Eq.~(\ref{recur1}). We will perform this numerical check for the parameter
choice $\alpha_1=\alpha_2=1$ which is relevant for the differential equation,
and for $D=2$ space-time dimensions. As mentioned in
Ref.~\cite{MullerStach:2011ru}, even though other dimensions are feasible, this
choice avoids singular contributions and serves for the simplest
integrand. The equation to be checked is
\begin{equation}\label{recur12}
2m_1^2S(2,1)+(p^2+m_1^2+m_2^2)S(1,2)=S(1,1)+S(0,2).
\end{equation}
Starting from
\begin{equation}
S(1,1)=\int\frac{d^2k}{(2\pi)^2}\frac1{(k^2+m_1^2)
  \left((p-k)^2+m_2^2\right)}
  =\frac1{2\pi}\int_0^\infty J_0(px)K_0(m_1x)K_0(m_2x)dx,
\end{equation}
the integrals $S(\alpha_1,\alpha_2)$ with higher (integer) values of
$\alpha_i$ can be obtained as partial derivatives with respect to the masses,
\begin{eqnarray}
\frac{-1}{2m_1}\frac\partial{\partial m_1}S(1,1)
  &=&-\frac\partial{\partial m_1^2}S(1,1)
  \ =\ -\frac\partial{\partial m_1^2}\int\frac{d^2k}{(2\pi)^2}
  \frac1{(k^2+m_1^2)\left((p-k)^2+m_2^2\right)}\nonumber\\
  &=&\int\frac{d^k}{(2\pi)^2}\frac1{(k^2+m_1^2)^2
  \left((p-k)^2+m_2^2\right)}\ =\ S(2,1)\qquad
\end{eqnarray}
and accordingly
\begin{equation}
\frac{-1}{2m_2}\frac\partial{\partial m_2}S(1,1)\ =\ S(1,2).
\end{equation}
In addition one has
\begin{eqnarray}
S(2,0)&=&\int\frac{d^2k}{(2\pi)^2}\frac1{(k^2+m_1^2)^2}
  \ =\ \frac1{4\pi m_1^2},\nonumber\\
S(0,2)&=&\int\frac{d^2k}{(2\pi)^2}\frac1{\left((p-k)^2+m_2^2\right)^2}
  \ =\ \frac1{4\pi m_2^2}.
\end{eqnarray}
The derivative can be expressed by $K'_0(z)=-K_1(z)$. The higher order 
integrals in the
configuration space representation are given by
\begin{eqnarray}
S(2,1)&=&\frac{-1}{2m_1}\frac\partial{\partial m_1}S(1,1)
  \ =\ \frac1{4\pi m_1^2}\int_0^\infty J_0(px)(m_1x)K_1(m_1x)K_0(m_2x)x\,dx,
  \qquad\nonumber\\
S(1,2)&=&\frac{-1}{2m_2}\frac\partial{\partial m_2}S(1,1)
  \ =\ \frac1{4\pi m_2^2}\int_0^\infty J_0(px)K_0(m_1x)(m_2x)K_1(m_2x)x\,dx.
\end{eqnarray}
Therefore, Eq.~(\ref{recur12}) in the configuration space representation reads
\begin{eqnarray}\label{recur12x}
1&=&\int_0^\infty J_0(px)\Bigg[2(m_2x)^2(m_1x)K_1(m_1x)K_0(m_2x)
  \nonumber\\&&\strut\qquad
  +((px)^2+(m_1x)^2+(m_2x)^2)K_0(m_1x)(m_2x)K_1(m_2x)\nonumber\\&&\strut\qquad
  -2(m_2x)^2K_0(m_1x)K_0(m_2x)\Bigg]\frac{dx}x.\qquad
\end{eqnarray}
We were able to check this equation numerically for different values of $p$
as function of $m_1$ and $m_2$. The 3D-plot in MATHEMATICA shows stochastic
fluctuations around the exprected value of $1$ of the order of $10^{-9}$.

Euler's theorem of homogeneous functions leads to the differential equation
\begin{equation}\label{euler}
\left(p^2\frac\partial{\partial p^2}+m_1^2\frac\partial{\partial m_1^2}
 +m_2^2\frac\partial{\partial m_2^2}+1\right)S(1,1)=0.
\end{equation}
Because of $J'_0(z)=-J_1(z)$, Eq.~(\ref{euler}) can be translated to
\begin{eqnarray}
0&=&\int_0^\infty(px)^2\Bigg[(px)J_1(px)K_0(m_1x)K_0(m_2x)
  +J_0(px)(m_1x)K_1(m_1x)K_0(m_2x)\nonumber\\&&\strut\qquad
  +J_0(px)K_0(m_1x)(m_2x)K_1(m_2x)
  -2J_0(px)K_0(m_1x)K_0(m_2x)\Bigg]\frac{dx}x\qquad
\end{eqnarray}
We checked on the latter relation with an even better precision of the order 
of $10^{-13}$.

The differential equation
\begin{equation}\label{remiddi1}
\left(p^2+(m_1+m_2)^2\right)\left(p^2+(m_1-m_2)^2\right)
  \frac\partial{\partial p^2}S(1,1)=-(p^2+m_1^2+m_2^2)S(1,1)
\end{equation}
in Ref.~\cite{Remiddi:1997ny} is obtained by inserting the recurrence relations
into Euler's differential equation~(\ref{euler}). Using the configuration space
representation,
Eq.~(\ref{remiddi1}) reads
\begin{eqnarray}
1&=&\int_0^\infty\Bigg[\frac{-1}{2(px)^2}\left((px)^2+(m_1x+m_2x)^2\right)
  \left((px)^2+(m_1x-m_2x)^2\right)(px)J_1(px)
  \nonumber\\&&\strut\qquad
  +\left((px)^2+(m_1x)^2+(m_2x)^2\right)J_0(px)\Bigg]K_0(m_1x)K_0(m_2x)
  \frac{dx}x.\qquad
\end{eqnarray}
This equation could be checked with a precision of the order of $10^{-8}$

While Euler's differential equation can be derived from general principles also
for the configuration space representation, the recurrence relations can be
derived only via the momentum space representation. If one does not use this
technique, it remains unclear why such integral identities exist for general
parameters $p$, $m_1$ and $m_2$. In order to check whether one can derive 
further relations
by using integral identities in configuration space,
we have used the general ansatz
\begin{eqnarray}
1&=&\int_0^\infty J_0(px)\Bigg[A_0(px,m_1x,m_2x)K_0(m_1x)K_0(m_2x)
  \nonumber\\&&\strut\qquad
  +A_1(px,m_1x,m_2x)(m_1x)K_1(m_1x)K_0(m_2x)\nonumber\\&&\strut\qquad
  +A_2(px,m_1x,m_2x)K_0(m_1x)(m_2x)K_1(m_2x)\Bigg]\frac{dx}x
\end{eqnarray}
where $A_i(p,m_1,m_2)=a_{i0}p^2+a_{i1}m_1^2+a_{i2}m_2^2$ ($i=0,1,2$). By
chosing random values for $p$, $m_1$ and $m_2$, and solving the resulting
system of equations, one obtains
\begin{eqnarray}
a_{00}=0,&& a_{01}=-2a,\qquad a_{02}=-2+2a,\nonumber\\[7pt]
a_{10}=a,&& a_{11}=a,\qquad a_{12}=2-a,\nonumber\\[7pt]
a_{20}=1-a,&&a_{21}=1+a,\qquad a_{22}=1-a,
\end{eqnarray}
where $a$ is an arbitrary parameter. This leads to Eq.~(\ref{recur12x}) and
\begin{eqnarray}
0&=&\int_0^\infty J_0(px)\Bigg[2\left((m_2x)^2-(m_1x)^2\right)
  K_0(m_1x)K_0(m_2x)\nonumber\\&&\strut\qquad
  +\left((px)^2+(m_1x)^2-(m_2x)^2\right)(m_1x)K_1(m_1x)K_0(m_2x)
  \nonumber\\&&\strut\qquad
  -\left((px)^2-(m_1x)^2+(m_2x)^2\right)K_0(m_1x)(m_2x)K_1(m_2x)\Bigg]
  \frac{dx}x\qquad
\end{eqnarray}
which is the difference of Eq.~(\ref{recur12x}) and the same equation with
$m_1$ and $m_2$ interchanged. We conclude that no more recurrence relations
can be found that go beyond Eq.~(\ref{recur12x}).

\section{The two-loop case with equal masses}
The differential equation for the two-loop degenerate sunrise diagram has
been given in Ref.~\cite{Laporta:2004rb}. It reads
\begin{eqnarray}\label{remiddi2}
\lefteqn{\Bigg(2p^2(p^2+m^2)(p^2+9m^2)\pfrac{d}{dp^2}^2
  +\left(3(4-D)p^4+10(6-D)m^2p^2+9Dm^4\right)\frac{d}{dp^2}}\nonumber\\&&\strut
  +(D-3)\left((D-4)p^2-(D+4)m^2\right)\Bigg)S(1,1,1)=\frac3{(D-4)^2\pi^2},
  \qquad\qquad
\end{eqnarray}
which simplifies to
\begin{equation}\label{remiddi22}
\Bigg(p^2(p^2+m^2)(p^2+9m^2)\pfrac{d}{dp^2}^2
  +(3p^4+20m^2p^2+9m^4)\frac{d}{dp^2}
  +(p^2+3m^2)\Bigg)S(1,1,1)=\frac3{8\pi^2}
\end{equation}
in $D=2$ space-time dimensions.
We write $S(1,1,1)$ in a form which is easily adapted to the 
non-degenerate mass case to be discussed later on. One has
\begin{equation}\label{S111}
S(1,1,1)
=\frac1{(2\pi)^2}\int_0^\infty J_0(px)K_0(m_1x)K_0(m_2x)K_0(m_3x)x\,dx\,.
\end{equation}
Differentiation of Eq.~(\ref{S111}) gives
\begin{eqnarray}\label{S111p}
\lefteqn{\frac{d}{dp^2}S(1,1,1)\ =\ \frac1{2p}\frac{d}{dp}S(1,1,1)}\nonumber\\
  &=&\frac{-1}{2(2\pi)^2p^2}\int_0^\infty(px)J_1(px)K_0(m_1x)K_0(m_2x)K_0(m_3x)
  x\,dx
\end{eqnarray}
and
\begin{eqnarray}\label{S111pp}
\pfrac{d}{dp^2}^2S(1,1,1)&=&\frac1{4(2\pi)^2p^4}\int_0^\infty
  \Bigg[(px)J_1(px)+\frac12(px)^2\left(J_2(px)-J_0(px)\right)\Bigg]\times
  \strut\nonumber\\&&\strut\times
  K_0(m_1x)K_0(m_2x)K_0(m_3x)x\,dx.
\end{eqnarray}
Returning to the degenerate mass case the differential 
equation~(\ref{remiddi22}) can be translated to
\begin{eqnarray}
\frac32&=&\frac1{4p^2}\int_0^\infty\Bigg[(p^2+m^2)(p^2+9m^2)
  \left((px)J_1(px)+\frac12(px)^2\left(J_2(px)-J_0(px)\right)\right)
  \nonumber\\&&\strut
  -2(3p^4+20m^2p^2+9m^4)(px)J_1(px)+4p^2(p^2+3m^2)J_0(px)\Bigg]K_0(mx)^3x\,dx.
  \qquad
\end{eqnarray}
where $J''_0(z)=-J'_1(z)=(J_2(z)-J_0(z))/2$ is used. Because the result
contains the second derivative of the Bessel function, one can use Bessel's
differential equation
\begin{equation}
z^2J''_\lambda(z)+zJ'_\lambda(z)+(z^2-\lambda^2)J_\lambda(z)=0
\end{equation}
for $\lambda=0$ to compactify the result,
\begin{eqnarray}
\frac32&=&\frac1{4p^2}\int_0^\infty\Bigg[\left(4p^2(p^2+3m^2)
  -(p^2+m^2)(p^2+9m^2)(px)^2\right)J_0(px)\nonumber\\&&\strut\qquad
  -2p^2(p^2+5m^2)J_1(px)\Bigg]K_0(mx)^3x\,dx.
\end{eqnarray}
These results have been checked with a precision of the order of $10^{-8}$.

\section{The two-loop case with arbitrary masses}
The final section of this paper is devoted to the second order differential
equation, derived for the two-loop sunrise diagram with arbitrary masses in
Ref.~\cite{MullerStach:2011ru}. After adjusting the normalization, the
differential equation can be written as
\begin{equation}
\Bigg[p_0(-p^2)\pfrac{d}{dp^2}^2
  +p_1(-p^2)\frac{d}{dp^2}+p_2(-p^2)\Bigg]S(1,1,1)
  =\frac{p_3(-p^2)}{4(2\pi)^2},
\end{equation}
where the coefficients $p_i(t)$ ($i=0,1,2,3$) are given by
\begin{eqnarray}
p_0(t)&=&t\left(t-(m_1+m_2+m_3)^2\right)\left(t-(-m_1+m_2+m_3)^2\right)
  \left(t-(m_1-m_2+m_3)^2\right)\times\strut\nonumber\\&&\strut\qquad\times
  \left(t-(m_1+m_2-m_3)^2\right)\left(3t^2-2M_{100}t-M_{200}+2M_{110}\right),
  \qquad\\[12pt]
p_1(t)&=&9t^6-32M_{100}t^5+(37M_{200}+70M_{110})t^4
  -(8M_{300}+56M_{210}+144M_{111})t^3\nonumber\\&&\strut
  -(13M_{400}-36M_{310}+46M_{220}-124M_{211})t^2\nonumber\\&&\strut
  -(-8M_{500}+24M_{410}-16M_{320}-96M_{311}+144M_{221})t\nonumber\\&&\strut
  -(M_{600}-6M_{510}+15M_{420}-20M_{330}+18M_{411}-12M_{321}-6M_{222}),
  \qquad\\[12pt]
p_2(t)&=&3t^5-7M_{100}t^4+(2M_{200}+16M_{110})t^3
  +(6M_{300}-14M_{210})t^2\nonumber\\&&\strut
  -(5M_{400}-8M_{310}+6M_{220}-8M_{211})t\nonumber\\&&\strut
  +(M_{500}-3M_{410}+2M_{320}+8M_{311}-10M_{221}),\qquad\\[12pt]
p_3(t)&=&-18t^4+24M_{100}t^3+(4M_{200}-40M_{110})t^2
  +(-8M_{300}+8M_{210}+48M_{111})t\nonumber\\&&
  +(-2M_{400}+8M_{310}-12M_{220}-8M_{211})
  +2c(t,m_1,m_2,m_3)\ln(m_1^2/\mu^2)\nonumber\\&&\strut
  +2c(t,m_2,m_3,m_1)\ln(m_2^2/\mu^2)
  +2c(t,m_3,m_1,m_2)\ln(m_3^2/\mu^2),
\end{eqnarray}
and where
\begin{equation}
M_{\lambda_1\lambda_2\lambda_3}=\sum_\sigma(m_1^2)^{\sigma(\lambda_1)}
  (m_2^2)^{\sigma(\lambda_2)}(m_3^2)^{\sigma(\lambda_3)}
\end{equation}
are monomial symmetric polynomials in $m_1^2$, $m_2^2$ and $m_3^2$ and where
\begin{eqnarray}
\lefteqn{c(t,m_1,m_2,m_3)\ =\ \left(-2m_1^2+m_2^2+m_3^2\right)t^3}
  \nonumber\\&&\strut
  +\left(6m_1^4-3m_2^4-3m_3^4-7m_1^2m_2^2-7m_1^2m_3^2+14m_2^2m_3^2\right)t^2
  \nonumber\\&&\strut
  +\Big(-6m_1^6+3m_2^6+3m_3^6+11m_1^4m_2^2+11m_1^4m_3^2
  \nonumber\\&&\strut\qquad\qquad
  -8m_1^2m_2^4-8m_1^2m_3^4-3m_2^4m_3^2-3m_2^2m_3^4\Big)t\nonumber\\&&\strut
  +\Big(2m_1^8-m_2^8-m_3^8-5m_1^6m_2^2-5m_1^6m_3^2+m_1^2m_2^6+m_1^2m_3^6
  +4m_2^6m_3^2+4m_2^2m_3^6\nonumber\\&&\strut\qquad
  +3m_1^4m_2^4+3m_1^4m_3^4-6m_2^4m_3^4+2m_1^4m_2^2m_3^2-m_1^2m_2^4m_3^2
  -m_1^2m_2^2m_3^4)\qquad
\end{eqnarray}
(for details, cf.\ Ref.~\cite{MullerStach:2011ru}). Using Eqs.~(\ref{S111}),
(\ref{S111p}) and~(\ref{S111pp}), one obtains
\begin{eqnarray}
p_3(-p^2)&=&\int_0^\infty\Bigg[\frac{p_0(-p^2)}{p^4}
  \left((px)J_1(px)+\frac12(px)^2\left(J_2(px)-J_0(px)\right)\right)
  \nonumber\\&&\strut
  -2\frac{p_1(-p^2)}{p^2}(px)J_1(px)+4p_2(-p^2)J_0(px)\Bigg]
  K_0(m_1x)K_0(m_2x)K_0(m_3x)x\,dx\nonumber\\
  &=&\int_0^\infty\Bigg[\left(4p_2(-p^2)-\frac{p_0(-p^2)}{p^4}(px)^2\right)
  J_0(px)\nonumber\\&&\strut
  -2\frac{p_1^*(-p^2)}{p^2}(px)J_1(px)\Bigg]
  K_0(m_1x)K_0(m_2x)K_0(m_3x)x\,dx\qquad
\end{eqnarray}
where $p_1^*(t)=p_1(t)+p_0(t)/t$. Using different values for $p$ and $m_3$,
in terms of $m_1$ and $m_2$ we obtain a 3D-plot with MATHEMATICA which shows
again stochastic fluctuations of the order of $10^{-4}$. In the course of
our numerical checks we were able to identify two typos in the coefficients 
of $c(t,m_1,m_2,m_3)$ in the preprint version of Ref.~\cite{MullerStach:2011ru}
which we have corrected.

\section{Conclusions}
Using configuration space techniques, we were able to check numerically the
differential equations for sunrise-type diagrams found in the literature.
The precision of our numerical test is still quite moderate, but gives
sufficient confidence in the validity of the differential equations derived
by other means. For example, the introduction of artificial ``typos'' in the 
coefficients of the differential equations are easily discovered.
More rigorous tests would require the use of more stable integration routines
than those provided by MATHEMATICA. For the future we hope to find independent
routes to discover further relations between Bessel moments which may lead to
generalizations of the present findings to cases involving three-loop or even 
higher order sunrise-type diagrams.

\subsection*{Acknowledgments}
We want to thank Stefan Weinzierl and Anatoly Kotikov for helpful discussions
and Volodya Smirnov for encouragement. S.G.\ acknowledges the support by the
Estonian target financed Project No.~0180056s09, by the Estonian Science
Foundation under grant No.~8769, and by the Deutsche Forschungsgemeinschaft
(DFG) under No.~KO~1069/14-1. A.A.P.\ acknowledges partial support by
the RFFI grant No.~11-01-00182-a.

\end{document}